\pdfoutput=1
\documentclass[]{article}  
\usepackage{url}
\usepackage{graphicx}
\usepackage{amsfonts}
\usepackage{amssymb}
\usepackage{latexsym}

\newcommand{\hide}[1]{}

                            {\makeatletter
                           \gdef\xxxmark{%
    \expandafter\ifx\csname @mpargs\endcsname\relax 
\expandafter\ifx\csname @captype\endcsname\relax 
   \marginpar{xxx}
                                \else
                     xxx 
                                 \fi
                                \else
                     xxx 
                                 \fi}
              \gdef\xxx{\@ifnextchar[\xxx@lab\xxx@nolab}
      \long\gdef\xxx@lab[#1]#2{{\bf [\xxxmark #2 ---{\sc #1}]}}
             \long\gdef\xxx@nolab#1{{\bf [\xxxmark #1]}}
\gdef\turnoffxxx{\long\gdef\xxx@lab[##1]##2{}\long\gdef\xxx@nolab##1{}}%
                                  }

\newcommand{\squeezelist}{\setlength{\itemsep}{0pt}}

\newcommand{\ABox}{
\raisebox{3pt}{\framebox[6pt]{\rule{6pt}{0pt}}}
}
\newenvironment{proof}{{\bf Proof:}}{\hfill\ABox}

\newtheorem{theorem}{{\bf Theorem}}

\newtheorem{lemma}[theorem]{Lemma}

\newtheorem{property}[theorem]{Property}

\newcommand{\lemlab}[1]{\label{lemma:#1}}

\newcommand{\figlab}[1]{\label{fig:#1}}

\newcommand{\eqref}[1]{\ref{eq:#1}}
\newcommand{\figref}[1]{\ref{fig:#1}}

\begin{document}

\title{A Class of Convex Polyhedra\\
with Few Edge Unfoldings}

\author{
Alex Benton\thanks{DAMTP, Centre for Mathematical Science, Cambridge University, Cambridge CB3 0WA, UK.
\protect\url{A.Benton@damtp.cam.ac.uk}.}
\and
Joseph O'Rourke\thanks{Dept. Comput. Sci., Smith College, Northampton, MA 01063, USA.
\protect\url{orourke@cs.smith.edu}.}
}

\maketitle
\begin{abstract}
We construct a sequence of convex polyhedra on $n$ vertices
with the property that, as $n {\rightarrow} \infty$,
the fraction of its edge unfoldings that avoid overlap
approaches 0,
and so the fraction that overlap approaches 1.
Nevertheless, each does have (several) nonoverlapping
edge unfoldings.
\end{abstract}

\section{Introduction}
An \emph{edge unfolding} of a polyhedron is a cutting of the surface
along its edges that unfolds the surface to a single, nonoverlapping piece
in the plane.
It has long been an open question of whether or not every convex
polyhedron has an edge unfolding.\footnote{
   \url{http://cs.smith.edu/~orourke/TOPP/P9.html#Problem.9}
}
See~\cite[Chap.~22]{do-gfalop-07} for background and the current status
of this problem.

An early empirical investigation of this question led to the conjecture
that a random edge unfolding of a random convex polyhedron 
of $n$ vertices leads
to overlap with probability 1 as $n {\rightarrow} \infty$,~\cite{so-cru-87},\footnote{
   Data summarized in~\cite[p.~315]{do-gfalop-07}.
}
under any reasonable definition of ``random.''
It is easy to see that the cuts must form a spanning tree of the
polyhedron vertices.
It is known that there are
$2^{\Omega(\sqrt{F})}$
cut trees for a polyhedron of $F$ faces.
So the conjecture says that ``most'' of the exponentially many
cut trees lead to overlap.
Of course, even if most unfoldings overlap in this sense,
this is entirely compatible with the hypothesis that there always
exists at least one non-overlapping unfolding.

No progress has been made on this random-unfolding conjecture
(as far as we know), but Lucier~\cite{l-losuc-06} was able
to disprove several unfolding conjectures by carefully arranged polyhedra
that force what he calls \emph{2-local overlap}.\footnote{
   The faces that overlap are incident to the endpoints of a common edge
   in the unfolding.
}
Although not all our overlaps are 2-local, they are $k$-local
(in Lucier's notation)
for some small $k$, so our work can be viewed as following the spirit
of his investigations.

In this note we construct an infinite sequence of convex polyhedra
with the property that most of its unfoldings overlap,
in the sense that, as $n {\rightarrow} \infty$,
the number of its edge unfoldings that overlap
approaches 1.

\section{Banded Hexagon}
The construction is based on a particular example from~\cite{o-bupc-07},
which showed that it is impossible to extend 
band unfoldings to obtain edge unfoldings of prismatoids.
The details of the motivation for that work are not relevant here,
but we employ its central construction, which we now describe.

Consider a hexagon
formed by replacing each side of an equilateral triangle with two
nearly collinear edges.
The hexagon is then surrounded by a band of six identical quadrilaterals,
forming a slight convexity at all edges.
See Fig.~\figref{banded_hex_3D}.
\begin{figure}[htbp]
\centering
\includegraphics[width=0.75\linewidth]{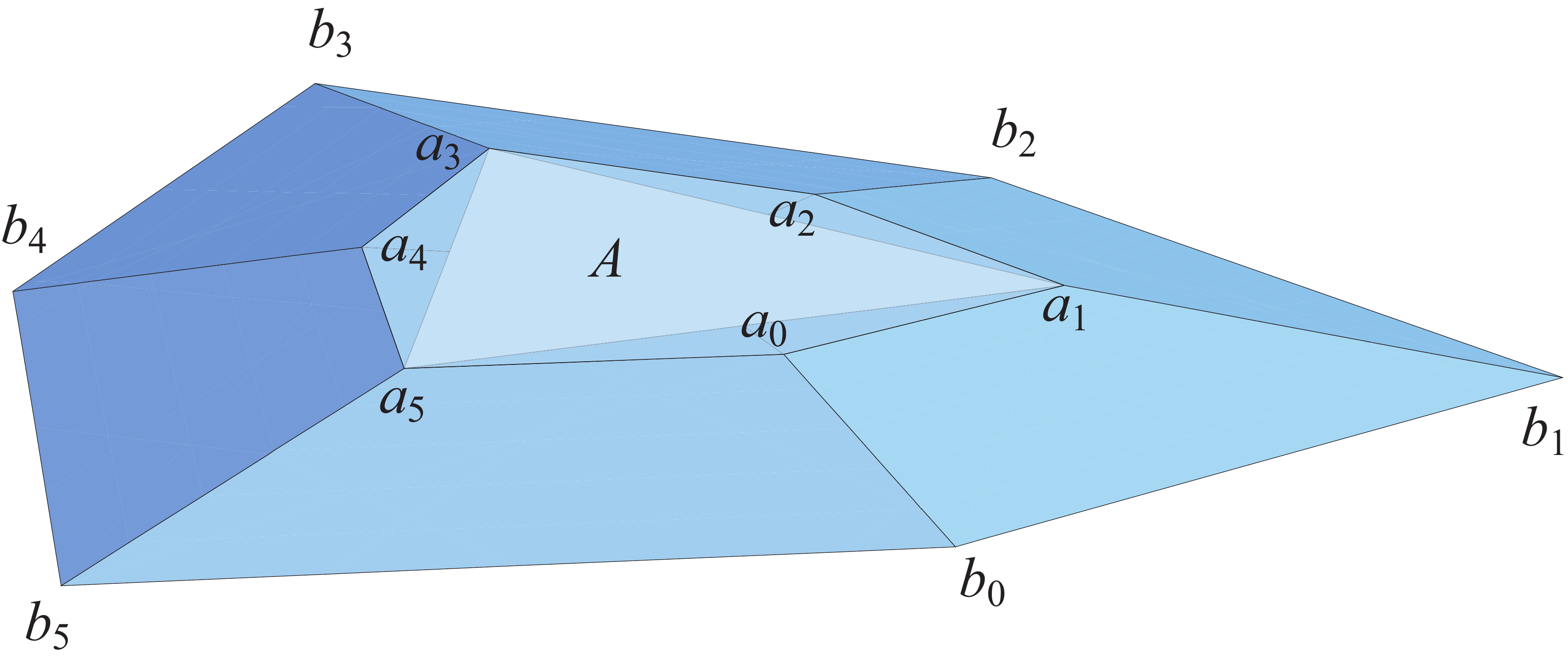}
\caption{Banded hexagon from~\protect\cite{o-bupc-07}.}
\figlab{banded_hex_3D}
\end{figure}
The six vertices of the hexagon $A$ are $(a_0,\ldots,a_5)$,
and each is connected to its counterpart $b_i$ on the outer rim
of the band.
The slight convexity means that the curvature at the $a_i$
vertices is small.
Cutting and flattening a vertex opens
it by an amount equal to the curvature.

The key property of this \emph{banded hexagon}
is as follows.
\begin{property}[Hexagon Overlap]
If only one band edge $a_i b_i$ is cut (as part of the cut tree),
so that the six faces of the band remain connected together,
and all but one of the hexagon edges $a_i a_{i+1}$ are cut,
then the unfolding overlaps.
\end{property}

Fig.~\figref{hexa_cex}(a-c) illustrates the opening at $a_3$,
and (d-f) the opening at $a_0$.  The other possibilities
are symmetric.
\begin{figure}[htbp]
\centering
\includegraphics[width=\linewidth]{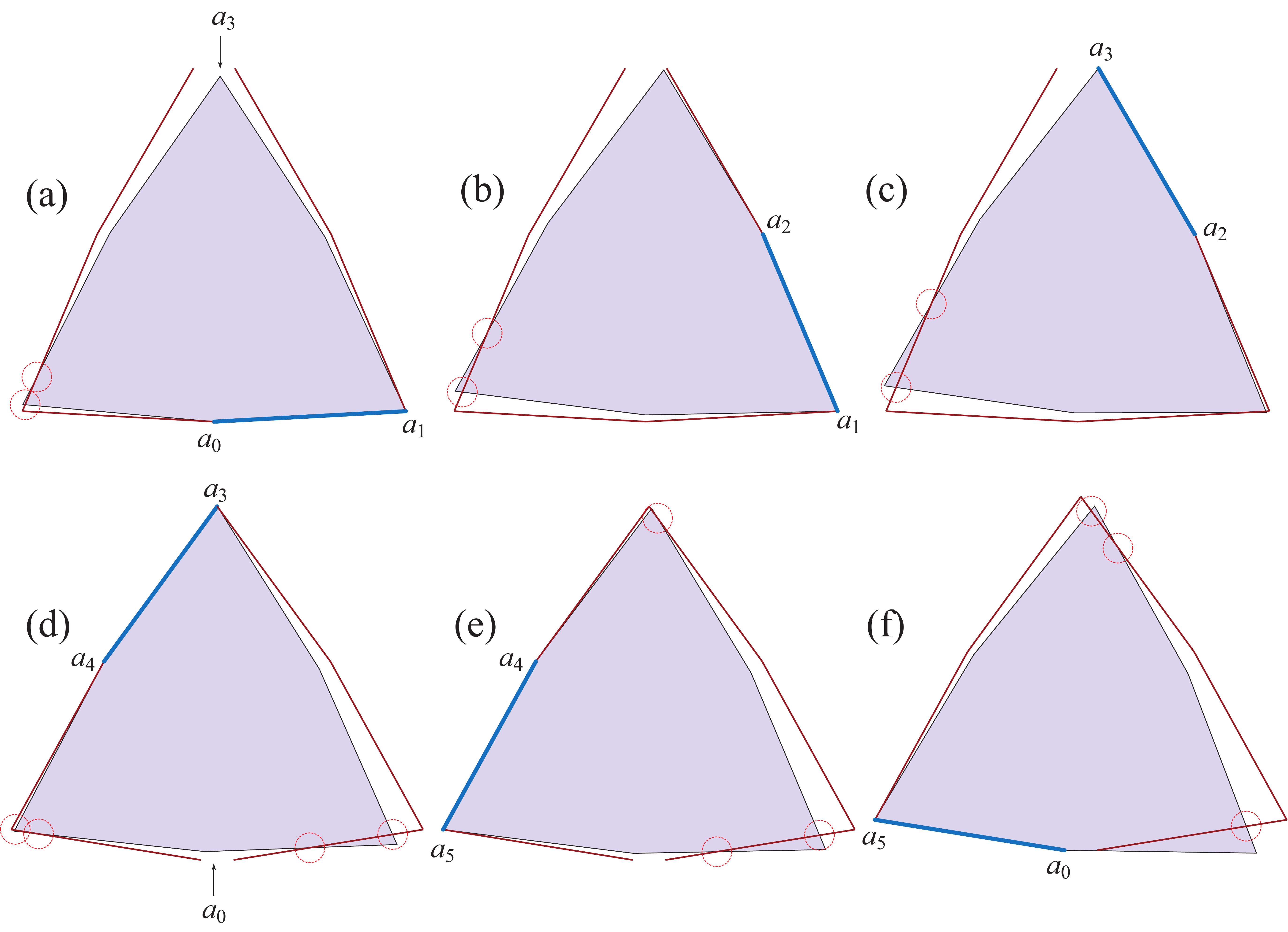}
\caption{Placements of $A$ when $a_3$ is cut
(top row) and when $a_0$ is cut (bottom row).
The attachment edge of the band to $A$ is blue.
Circles indicate overlap.
The band lies outside the red rim.
[Fig.~3 in~\protect\cite{o-bupc-07}.]
}
\figlab{hexa_cex}
\end{figure}

\section{Banded Geodesic Domes}
For the purposes of~\cite{o-bupc-07}, the band quadrilaterals were
chosen to be trapezoids.  However, that is not an essential property,
and we modify the construction here so that the quadrilaterals
remain congruent but are no longer trapezoids.
The Hexagon Overlap property only relies on small curvature at the $a_i$,
and the hexagon $A$ having three acute angles (at $\{a_1,a_3,a_5\}$)
interspersed with
three nearly $\pi$-angles (at $\{a_0,a_2,a_4\}$).
See ahead to Fig.~\figref{hex_dual}.

With this flexibility,
it is possible to glue together copies of the banded hexagon
construction onto a triangulated surface.
We choose to use ``geodesic domes'' as our base polyhedron,
a repeated meshing starting with the icosahedron
that has nearly equilateral faces.
Fig.~\figref{geodesic_domes} illustrates the first four \emph{levels}
of the geodesic dome construction, with each triangle face
replaced by a banded hexagon.
Let $P_L$ be the banded geodesic dome refined to level $L$.
Level $L=0$ is based on the icosahedron.
Level $L=1$ partitions each face 
of the icosahedron into four equilateral triangles,
and projects to the circumscribing sphere.  And so on.
The number of faces, edges, and vertices of the completed construction
for $P_L$ are:

\begin{eqnarray*}
F & = & 140 \cdot 4^L\\
E & = & 300 \cdot 4^L\\
n = V & = & 160 \cdot 4^L
\end{eqnarray*}

We can drive $n {\rightarrow} \infty$ by choosing larger and larger values
of $L$.  At $L=3$, there are $n=10242$ vertices.

\begin{figure}[htbp]
\centering
\includegraphics[width=\linewidth]{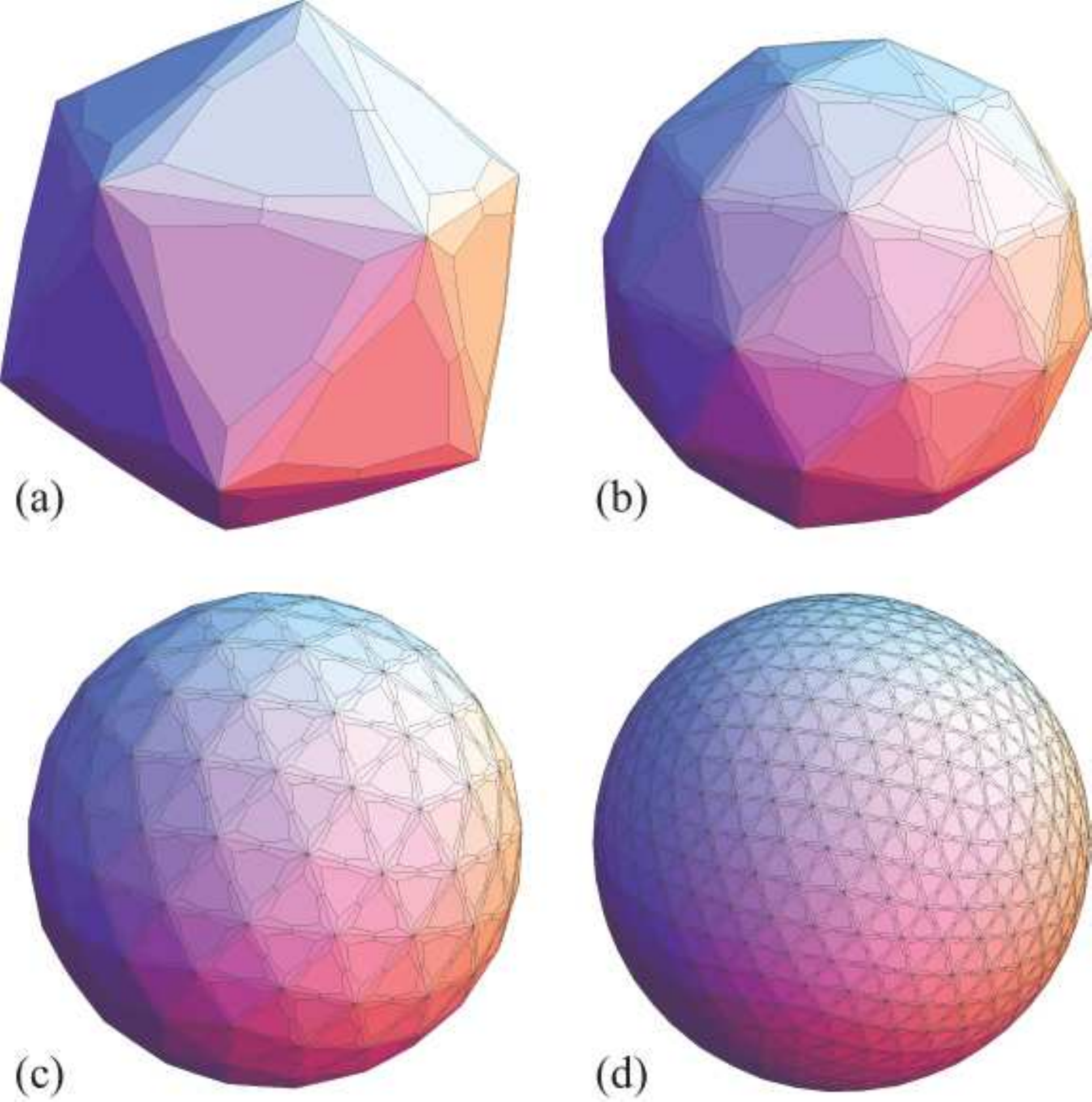}
\caption{Banded geodesic domes for levels $L=0,1,2,3$.
[Quality of this figure reduced to satisfy arXiv restrictions.]
}
\figlab{geodesic_domes}
\end{figure}

\section{Unfoldings}
Although the point of this note is that these banded geodesic domes
are in some sense difficult to edge-unfold,
in fact each of the four shown in Fig.~\figref{geodesic_domes}
can unfold without overlap.
Figs.~\figref{ebandunf0}-\figref{ebandunf3}
show unfoldings found by a yet-to-be-thwarted 
unfolding procedure
described in~\cite{b-f-08}.
Although we have not attempted to formally prove it,
it seems likely that banded geodesic domes for any $L$ can
be edge-unfolded similarly, roughly by following the geodesics.

\begin{figure}[thbp]
\begin{minipage}[b]{0.49\linewidth}
\centering
\includegraphics[width=\linewidth]{Figures/ebandunf0}
\caption{Edge unfolding of banded geodesic dome, $L=0$.}
\figlab{ebandunf0}
\end{minipage}%
\hspace{0.02\linewidth}%
\begin{minipage}[b]{0.49\linewidth}
\centering
\includegraphics[width=\linewidth]{Figures/ebandunf1}
\caption{Edge unfolding of banded geodesic dome, $L=1$.}
\figlab{ebandunf1}
\end{minipage}%
\end{figure}
\begin{figure}[htbp]
\begin{minipage}[b]{0.49\linewidth}
\centering
\includegraphics[width=\linewidth]{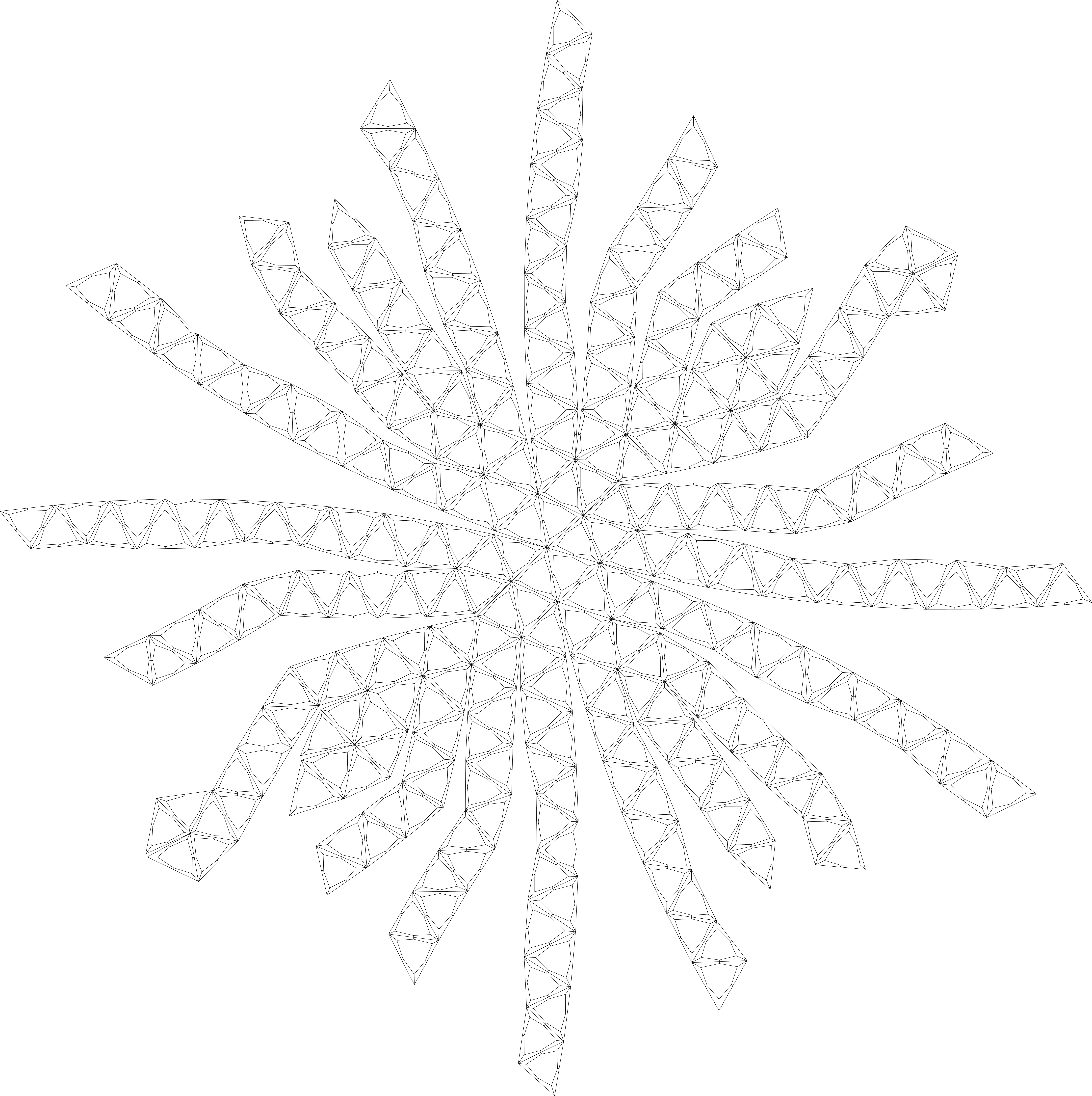}
\caption{Edge unfolding of banded geodesic dome, $L=2$.}
\figlab{ebandunf2}
\end{minipage}%
\hspace{0.02\linewidth}%
\begin{minipage}[b]{0.49\linewidth}
\centering
\includegraphics[width=\linewidth]{Figures/ebandunf3}
\caption{Edge unfolding of banded geodesic dome, $L=3$.
}
\figlab{ebandunf3}
\end{minipage}%
\end{figure}

All of these unfoldings have the property that each hexagon has
two or more band cuts incident to its vertices
(although these cuts are below the resolution of all
but  Fig.~\figref{ebandunf0}).  We see how this avoids
the Hexagon Overlap property in the next
section.

\section{Proof}
\paragraph{Overview.}
The proof has the following overall structure.
First we establish that at least a positive fraction $\rho > 0$ of
all cut trees that span a finite-sized connected region $C$ of the 
surface of $P_L$ satisfy the Hexagon Overlap property,
and so force unfolding overlap.
Thus, at most $(1-\rho)$ of those trees avoid overlap.
Then a cut tree that avoids overlap everywhere in the unfolding must avoid
local overlap in each of these regions.
Because the regions are a finite-size,
as $L{\rightarrow}\infty$, the number $k$ of regions
also gets arbitrarily large.  Thus
the fraction of trees that avoid overlap everywhere
is at most $(1-\rho)^k$, which goes to 0 as  $k{\rightarrow}\infty$.

\paragraph{Connection Tree.}
The cut tree $T$ is a spanning tree of the polyhedron vertices.
The dual \emph{connection} tree $T^{\triangle}$ is a spanning tree of the faces.
In $T^{\triangle}$, two face nodes are connected if the faces share an
uncut edge.
$T$ and $T^{\triangle}$ each uniquely determine the other.
In this section we reason mostly with $T^{\triangle}$.

\paragraph{One Hexagon.}
Focus on one hexagon $A$ of the polyhedron $P$.
Referring to Fig.~\figref{hex_dual},
let $e_i= a_i a_{i+1}$, and $u_i = a_i b_i$.
The conditions that lead to Hexagon Overlap
are: exactly one $e_i$ is not cut, and exactly one $u_i$ is cut.
In terms of the dual tree $T^{\triangle}$,
this means that the hexagon is a leaf node,
surrounded by a band path of length 5,
as in the figure.
Clearly there are $6^2$ such dual tree patterns leading
to Hexagon Overlap (6 choices for $e_i$ and 6 for $u_j$),
when one banded hexagon is considered in isolation.

\begin{figure}[htbp]
\centering
\includegraphics[width=0.5\linewidth]{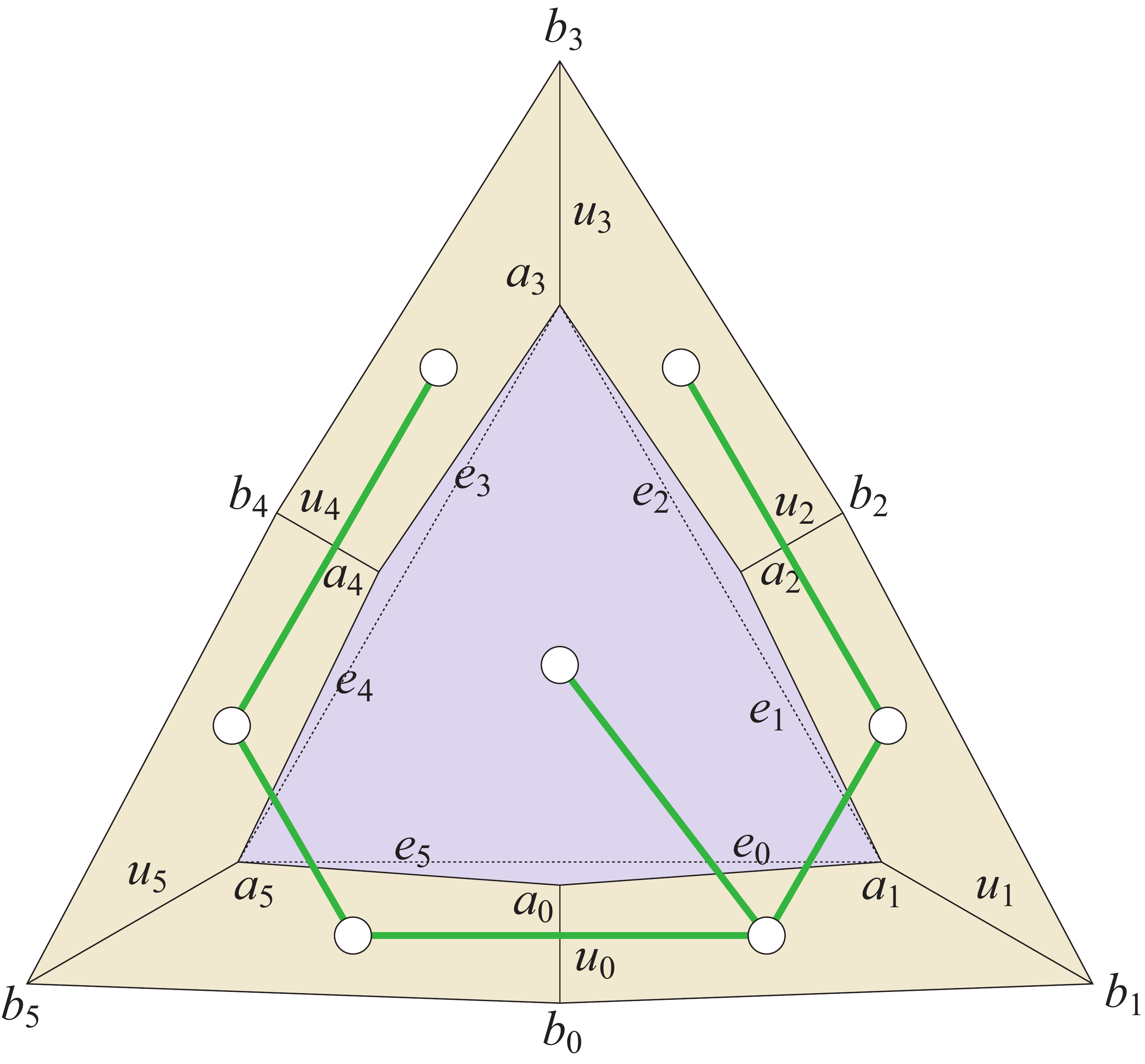}
\caption{$e_0$ is not cut and $u_3$ is cut.
All other $e_i$ are cut and all other $u_j$ are not cut.
Dual tree $T^{\triangle}$ is shown.
}
\figlab{hex_dual}
\end{figure}

\paragraph{Tiling Clusters.}
Now we consider a group of $16$ banded hexagons, which together
form a nearly equilaterial triangular cluster,
as shown in
Fig.~\figref{hexagon_tiling}.
\begin{figure}[htbp]
\centering
\includegraphics[width=0.75\linewidth]{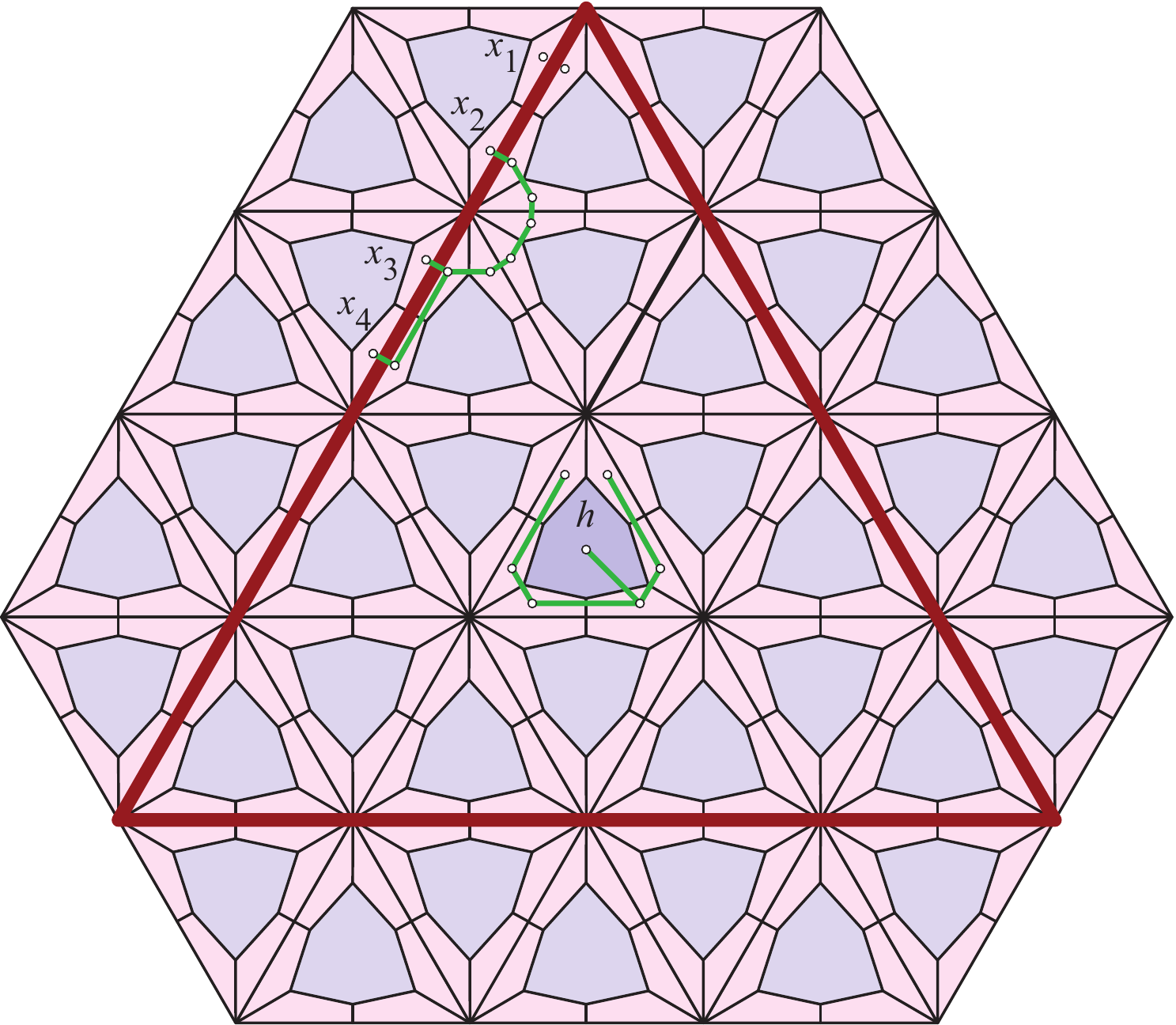}
\caption{$C$: $16$ banded hexagons, with central $h$.
$x_1, \ldots, x_{24}$: surrounding quadrilateral nodes.
}
\figlab{hexagon_tiling}
\end{figure}
Let $h$ be the central banded hexagon in a cluster $C$.
The choice of the size and shape of $C$ is somewhat arbitrary.
Our specific choice is motivated by two concerns:
(1)~The surface of $P_L$ is nearly an equilateral lattice tiling
of banded hexagons, and so can itself be tiled by copies of
the nearly equilateral $C$, for appropriate $L$.
(2)~The central $h$ is sufficiently ``buffered'' from the boundary of $C$,
in this case by the $15$ other banded hexagons of $C$,
for a counting argument to go through.
Both of these points will be revisited below.

\paragraph{Counting Overlapping Trees.}
We now argue that there are at least a positive fraction $\rho > 0$
of trees spanning $C$ that induce local overlap.

Let $T^{\triangle}$ be a dual spanning tree of $P$,
and denote by $G^{\triangle}$ the forest with all nodes in $C$ deleted.
There are in general many ways to complete $G^{\triangle}$ to
be a spanning tree of $P$.
The exact number of completions is difficult to count because
it depends on the structure of $G^{\triangle}$.
However, we can easily obtain a crude upper bound
as follows.
Let $E_C$ be the number of dual edges in $C$;
an explicit count shows that $E_C = 228$.
Any completion must either use or not use each dual edge in $C$.
Of course many of these ``bit patterns'' will not complete
$G^{\triangle}$ to a tree, or not to a spanning tree.
But every valid completion corresponds to one of these bit patterns.
Therefore, the total number of completions $m$ satisfies $m \le 2^{E_C}$.

Let $o$ be the number of completions of $G^{\triangle}$ that lead to
unfolding overlap.
Again it would be difficult to count $o$ exactly, 
but we know that the $36$ patterns leading to Hexagon Overlap in $h$
must be avoided, for each forces local overlap.
Moreover, because of the buffer around $h$ in $C$,
all of these $36$ patterns are part of some valid completion,
regardless of the structure of $T^{\triangle}$ outside $C$.
We justify this last claim below, but for now proceed with the
argument, assuming $o \ge 36$.

Let $\rho = o/m$ be the fraction of completions of $G^{\triangle}$ that
lead to overlap.  We have a lower bound on $o$
and an upper bound on $m$,
so together they provide a lower bound on the ratio $\rho$:
$$\rho \ge 36 / 2^{228} \approx 10^{-67} \;.$$
The exact value of this fraction $\rho$ is not relevant to the argument;
we only need that $\rho > 0$ so that $1-\rho < 1$.

\paragraph{Buffering.}
We return to the claim that $h$ is sufficiently buffered within $C$
so that for each tree that spans $C$, there are at least the
$36$ overlapping variants identified above.
First we explain why the more natural choice of $C=h$
does not suffice.
Suppose the forest $G^{\triangle}$ has the structure
that choosing an edge dual to $u_i$ within $h$ creates a cycle.
Then it is not a option to select this edge to complete
$G^{\triangle}$ to a tree.
If this occurs for two or more of the $u_i$, then the
Hexagon Overlap pattern of Fig.~\figref{hex_dual}
cannot occur within $h$.
Thus, the structure of $G^{\triangle}$ outside $C$ forces
avoidance of the Hexagon Overlap property inside $C$.
Thus, not every $C$ contains something to be avoided, so to speak.
We now show that our choice for $C$ provides sufficient buffering.

Let $x_1, x_2, \ldots, x_{24}$ be the $24$ quadrilateral nodes
surrounding and just outside $C$, each with a dual edge
that crosses into $C$.
Each can be viewed as the root of a tree in the forest $G^{\triangle}$.
We now show that the $36$ critical patterns are part of some completion
of $G^{\triangle}$ to a tree that spans $C$ and therefore all of $P_L$.
We first connect up all these trees in the forest into one tree via
connections through the quadrilaterals incident to the border of $C$.
One way to do this is to proceed sequentially from $x_1$ to $x_{23}$,
connecting $x_i$ to $x_{i+1}$ if their two subtrees are not
yet connected, but not making the connection if they already
are part of the growing connected component.
(For example, in  Fig.~\figref{hexagon_tiling}, perhaps $x_1$ does not
need to be connected to $x_2$, but $\{x_2, x_3, x_4 \}$ should receive connections.)
This connects all of $G^{\triangle}$ into a single tree
without employing any of the nodes of the central $h$.
For each of the $36$ overlap patterns for $h$, we are
free to connect up the remainder of $C$
into a spanning tree structure, which clearly can be done in many ways.
Therefore, for any tree that spans $P_L$ and $C$, there are
at least $36$ variants inside $C$ that overlap, and so $o \ge 36$.

We should remark that $C$ is larger than is needed for this argument
to go through (e.g., the three banded hexagons at the three
corners of $C$ are not needed), 
but it is easier to tile the surface if we agglomerate
into a nearly equilateral $C$.

\paragraph{Tiling $P_L$ with Clusters $C$.}
Within each original icosahedral face, the geodesic dome
partitioning creates a equilateral triangle tiling, which is
projected to the circumscribing sphere.
Each level increases the number of triangles by a factor of 4,
so every two-level increase multiplies by 16.
Thus, for even $L > 0$, we can tile each original icosahedral face with 
our 16-hexagon clusters.
This first applies for $L=2$, Fig.~\figref{geodesic_domes}(c).

\paragraph{Global Argument.}

Let $H = 20 \cdot 4^L$ be the number of hexagons in the polyhedron $P$.
We showed above that at most $1-\rho$ of the dual cut tree patterns 
inside a given cluster avoid overlap there
(for if we fall into the $\rho$ fraction, overlap is forced).

Imagine now constructing a complete tree $T^{\triangle}$
cluster-by-cluster in the tiling,
by choosing all the nodes and arcs in $T^{\triangle}$
that span one cluster $C$, before moving to the next cluster.
This is would be an odd way to build the tree,
but with appropriate foresight, any tree could be constructed
in this manner.
Selecting the subforest to span a particular $C$
leads us into the analysis of above:
no matter what the structure of $G^{\triangle}$ 
already fixed outside of $C$, there is a fraction $\rho$ of
subforests that must be avoided inside $C$.

In order to avoid overlap in the complete unfolding,
one of these overlap-avoiding patterns must be selected 
for each of the $\lfloor H/16 \rfloor$ clusters that tile the surface.
(We use the floor function here, but as noted above, choosing $L$ to be even
makes the tiling exact.)
Thus, the fraction of trees that avoid overlap within all clusters
simultaneously
is at most $(1-\rho)^{\lfloor H/16 \rfloor}$.

Finally, as $L {\rightarrow} \infty$, $H {\rightarrow} \infty$,
and the overlap-avoiding fraction of all unfoldings goes to 0,
while the overlap fraction goes to 1.
This is the main claim of this note.

\section{Empirical Data}
The argument above only establishes a (very) loose upper bound on the ratio
of the overlap-avoiding unfoldings to the total number of unfoldings.
Overlap can occur for other reasons, for example, by interactions
between non-adjacent faces of the polyhedron.  Our computation
is only concerned with avoiding a particular type of local overlap.
And the argument is very cautious; for example, $\rho$ is certainly much
larger than our minuscule lower bound.

The looseness of the argument is dramatically revealed by
empirical results.
Because the number of cut trees is so large, it is difficult
to obtain exact counts.  Instead we
generated random spanning cut trees, and checked each
for overlap in the resulting planar unfolding.\footnote{
   The software is described in~\cite{b-f-08}.
}
For the $L=0$ banded geodesic dome, 
our bound\footnote{
   Technically, the bound only applies to even $L > 0$,
   but we use it here just for a magnitude a comparison.
}
says that the overlapping-avoiding fraction is at most $1-10^{-67}$,
i.e., a random cut tree could almost always avoid overlap
according to our bound.
However, our simulations found only 11
non-overlapping unfoldings out of 5.5 million random cut trees,
for a ratio of about $2 \times 10^{-6}$,
i.e., overlap is almost never avoided, with 99.9998\% of unfoldings overlapping.
For higher values of $L$, no random cut tree led to non-overlap.
Even for a level-0 banded tetrahedron, only about 9\% of random unfoldings
were non-overlapping.
So the overlap-avoiding fraction of all unfoldings of banded
geodesic domes goes to zero
much, much faster than our crude analysis indicates.
Correspondingly, almost all unfoldings of these domes overlap.

Some understanding of this high frequency of overlap is
provided by the empirical observation that, in our random
unfoldings, about 70\% unfolded the seven faces of a banded hexagon 
connected together as a unit.  This fraction is stable and
apparently independent of $L$
(and therefore of $n$).\footnote{
    We have not attempted a theoretical explanation for this data.
}
And when a banded hexagon is unfolded
as a unit, the empirically observed frequency of local overlap
is about 50\%.  Thus, we would expect the fraction
$$ 1 - (1 - 0.7 \cdot 0.5) ^ H $$
of all unfoldings to overlap.
For $L=0$, $H=20$, this formula (using more accurate frequencies)
evaluates to 99.97\%.
This suggests that local overlap (within one banded hexagon unit)
accounts for the majority of overlaps, for counting
all overlaps only increases the frequency to 99.9998\%.

\section{Discussion}
The process of replacing each triangular face of a polyhedron
by a banded hexagon could be carried out on any triangulated polyhedron,
even if the faces are not nearly equilateral as in our geodesic domes.
We believe this always produces a polyhedron difficult to unfold
in the sense we have established here.
It may be that some ``base'' polyhedra will yield improvements
over the geodesic domes.
This remains to be explored.

Finally, through an independent argument that we will not detail,
we claim that at least $3/2^{17}$ of all
edge unfoldings of geodesic domes overlap, i.e., there is a fixed
fraction independent of $n$ that overlap.
Our experiments reported above indicate this is a significant underestimate
of the true fraction that overlap, but it is fraction that can
be proved.
Although this claim is in some sense weaker and less interesting
than the $n {\rightarrow} \infty$ result,
it
naturally raises the question of finding (infinite) classes of polyhedra
for which a larger fraction of all edge unfoldings can be proved to overlap,
i.e., classes more difficult to edge-unfold.

\paragraph{Acknowledgments.}
The second author thanks Erik Demaine for several enlightening
discussions on the topic of this paper.



\bibliographystyle{alpha}
\bibliography{/home/orourke/bib/geom/geom}
\end{document}